\documentclass[11pt,a4paper]{article}

\usepackage{amsmath,amssymb}
\usepackage{graphicx}
\usepackage{cite}
\usepackage{url}

\renewcommand\[{\left[}

\def\beq{\begin{equation}}
\def\eeq{\end{equation}}

\begin{document}

\numberwithin{equation}{section}
\title{{\normalsize  \mbox{}\hfill DCPT/13/04, IPPP/13/02}\\
\vspace{2.5cm}
\Large{\textbf{Emergence of the Electroweak Scale through the Higgs Portal}}}

\author{Christoph Englert$^1$, Joerg Jaeckel$^{2}$, Valentin V. Khoze$^1$, Michael Spannowsky$^{1,3}$\\[2ex]
  \small{\em $^1$Institute for Particle Physics Phenomenology, Department of Physics,} \\
  \small{\em South Road, Durham DH1 3LE, United Kingdom}\\[0.8ex]
  \small{\em $^2$Institut f\"ur theoretische Physik, Universit\"at Heidelberg,}\\
  \small{\em Philosophenweg 16, 69120 Heidelberg, Germany}\\[0.8ex]
  \small{\em $^3$Centre for Academic Practice, School of Education,}\\
   \small{\em  Leazes Road, Durham, DH1 1TA, United Kingdom }\\[0.8ex]
}

\date{}
\maketitle

\begin{abstract}
  \noindent
  Having discovered a candidate for the final piece of the Standard
  Model, the Higgs boson, the question remains why its vacuum
  expectation value and its mass are so much smaller than the Planck
  scale (or any other high scale of new physics).  One elegant
  solution was provided by Coleman and Weinberg, where all mass scales
  are generated from dimensionless coupling constants via dimensional
  transmutation.  However, the original Coleman-Weinberg scenario
  predicts a Higgs mass which is {\it too light}; it is parametrically
  suppressed compared to the mass of the vectors bosons, and hence is
  much lighter than the observed value. In this paper we argue that a
  mass scale, generated via the Coleman-Weinberg mechanism in a hidden
  sector and then transmitted to the Standard Model through a Higgs
  portal, can naturally explain the smallness of the electroweak scale
  compared to the UV cutoff scale, and at the same time be consistent
  with the observed value. We analyse the phenomenology of such a
  model in the context of present and future colliders and low energy
  measurements.
\end{abstract}
\thispagestyle{empty}
\setcounter{page}{0}

\newpage

\section{Introduction}\label{sec:intro}
The recent discovery of a particle which is likely to be the Higgs
boson~\cite{orig,ATLAS:2012gk,CMS:2012gu} with a mass of $\sim
125~{\rm GeV}$ concludes the quest to complete the particle spectrum
of the Standard Model.  However, the Standard Model itself leaves many
open questions. Most crucially, the question of the origin of the
electroweak scale remains unanswered.  Let us briefly consider the
Higgs potential in the Standard Model,
\begin{equation}
  \label{VH}
  V(H)\,=\, \mu^{2}_{\rm SM} \,H^{\dagger}H \,+
  \, \frac{\lambda_{\rm H}}{2}\,\left(H^{\dagger}H\right)^{2}\,,
\end{equation}
for the Higgs doublet $H$ which in the unitary gauge takes the form
$H^T(x)=\frac{1}{\sqrt{2}}(0, v+h(x))$. The minimum of the potential
occurs at $v^2 = - 2\mu^{2}_{\rm SM} / \lambda_{\rm H}$ for negative
$\mu^{2}_{\rm SM}$ and the mass of the Higgs boson $h$ is $m_h^2 =
\lambda_{\rm H} v^2$.

Choosing a value
\begin{equation}
  \label{musm}
  \mu^{2}_{\rm SM}\, =\, -\,\frac{1}{2}\,\lambda_{\rm H} \, v^2 \, 
  =\, -\,\frac{1}{2}\, m_h^2
\end{equation}
for the Higgs mass parameter $\mu^{2}_{\rm SM}$ in \eqref{VH}, an
expectation value $v\simeq 246~{\rm GeV}$ for the Higgs field and the
Higgs mass $m_h \sim 125~{\rm GeV}$ can be easily accommodated.
However, the Standard Model itself cannot explain the value of this
parameter and in particular its smallness compared to the UV cutoff,
$M_{UV}$ (which we take to be the scale of new physics in the UV where
the Standard Model breaks down as an effective theory, {\it e.g.} $M_{Pl}$).

In a seminal paper~\cite{Coleman:1973jx} Coleman and Weinberg showed
that in the absence of mass scales in the potential of a scalar field,
a mass scale is nevertheless generated via dimensional transmutation
from the running couplings, and indeed spontaneous symmetry breaking
does occur. A minimal self-consistent theory for this mechanism at
work is provided by massless scalar QED. This is a model with a
massless complex scalar field\footnote{We point out that we use the
    same normalisation as Coleman and Weinberg, treating the complex
    field $\phi=\phi_{1}+i\phi_{2}$ as two real scalar fields with kinetic term
    $\frac{1}{2}(\partial_{\mu}\phi_{1}\partial^{\mu}\phi_{1}+
    \partial_{\mu}\phi_{2}\partial^{\mu}\phi_{2})$.},
\begin{equation}
  V_{\rm cl}=\frac{\lambda_{\phi}}{4!}|\phi|^4\,,
\end{equation}
charged under a U(1) symmetry with gauge coupling $e_{\phi}$.
Starting from a classical potential and requiring that the
renormalised mass term for $\phi$ vanishes, the authors
of~\cite{Coleman:1973jx} find the 1-loop corrected potential
\begin{equation}
  \label{Veff1}
  V(\phi)\,=\, V_{\rm cl}+\Delta V_{\rm 1-loop}\,=
  \, \frac{\lambda_{\phi}}{4!}|\phi|^{4}\,+\,
  \left(\frac{5\lambda_{\phi}^2}{1152\pi^2}
    +\frac{3 e_{\phi}^{4}}{64\pi^2}\right)|\phi|^{4}
  \left[\log\left(\frac{|\phi|^{2}}{M^{2}}\right)
    -\frac{25}{6}\right]\,,
\end{equation}
where $M$ is the renormalisation scale.

The essential feature/requirement employed here is that the
\emph{renormalised mass} at the origin in the field space is kept at
zero,
\begin{equation}
  \label{eq:m0}
  m^2\, := \, V^{\prime\prime}(\phi)\bigg|_{\phi=0}=0 \, .
\end{equation}
In dimensional regularisation, which does not introduce any explicit
scale aside from the RG scale, entering the logarithmically running
couplings, this equation is satisfied automatically.\footnote{No
  power-like divergencies proportional to the cutoff scale appear
  in dimensional regularisation, and in theories like ours, which
  contain no explicit mass scales at the outset, no {\it finite}
  corrections to dimensionful quantities can appear either.}  In other
regularisation schemes such as {\it e.g.} the cutoff scheme, the zero on the
right hand side of \eqref{eq:m0} corresponds to an exact cancellation
of all the quadratically divergent parts between the bare mass squared
terms and the counterterms.

The consequence of \eqref{eq:m0} is that no explicit mass scales are
allowed in the effective potential of the theory, except the
renormalisation scale appearing in the logarithm.  This is the
manifestation of the scale invariance of the classical massless
theory; the scale invariance is broken only by the radiative
corrections which introduce only a logarithmic scale dependence.

Now returning to the effective potential in \eqref{Veff1}, at small
values of the field, the logarithm in the brackets always wins, giving
the potential a downward slope.  On the other hand, at values $|\phi|>
M$ the slope is always positive.  Accordingly we always have a minimum
of the potential at a value $|\phi|>0$.  Using this, one can remove
the renormalisation scale $M$ of the potential by renormalising at the
acquired vacuum expectation value (vev) $\langle|\phi|\rangle>0$.  Another
simplification arises from the fact that if we choose $e^{2}_{\phi}\ll
1$ the value of the $\phi$-self-coupling $\lambda_{\phi}$ squared at
the minimum of the effective potential is negligible compared to the
U(1) gauge coupling $e_{\phi}^2$, as shown in Eq.~\eqref{eq:rad}
below, and we can drop the first term in brackets \eqref{Veff1}. We
thus have
\begin{equation}
  \label{Veff2}
  V(\phi)\,=\,
  \frac{\lambda_{\phi}}{4!}|\phi|^{4}\,+\,
  \frac{3 e_{\phi}^{4}}{64\pi^2}|\phi|^{4}
  \left[\log\left(\frac{|\phi|^{2}}{\langle |\phi|^{2}\rangle}\right)
    -\frac{25}{6}\right]\,,
\end{equation}
The minimum of the effective potential is at
\begin{equation}
  \label{Veffprime}
  V'\,=\, \frac{1}{6}\left( \lambda_{\phi}-
    \frac{33}{8\pi^2}e^{4}_{\phi}\right)
  {\langle \phi\rangle}^3 \, =\, 0
\end{equation}
and the vev $\langle \phi\rangle$ is determined by the condition on
the couplings renormalised at the scale of the
vev~\cite{Coleman:1973jx},
\begin{equation}
  \label{eq:rad}
  \lambda_{\phi}(\langle|\phi|\rangle)
  =\frac{33}{8\pi^2}e^{4}_{\phi}(\langle |\phi|\rangle)\,.
\end{equation}
The effective potential in the vacuum reads
\begin{equation}
  \label{CWpotential}
  V(\phi)=\frac{3e^{4}_{\phi}}{64\pi^2}|\phi|^{4}
  \left[\log\left(\frac{|\phi|^{2}}{\langle |\phi|^{2}\rangle}\right)
    -\frac{1}{2}\right]\,.
\end{equation}

Since the couplings run only logarithmically the vev fixed by the
condition \eqref{eq:rad} depends exponentially on the coupling
constants.  In fact, in weakly coupled perturbation theory the vev is
naturally generated at the scale which is exponentially smaller than
the UV cutoff.  This can be illustrated by solving the leading-order
RG-running equation for the coupling $e_{\phi}$,
\begin{equation}
  \label{run1}
  \frac{d e_{\phi} }{dt} \,= \, \frac{ e_{\phi}^3}{48 \pi^2} 
  \ , \qquad {\rm where} \qquad
  t= \log (M/\Lambda_{UV})\,.
\end{equation}
Upon integration and setting the RG scale $M=\langle |\phi|\rangle$ we find
\begin{equation}
  \label{run2}
  \langle |\phi|\rangle \,=\,\Lambda_{UV} 
  \exp\left[-24\pi^2\left(\frac{1}{e^{2}_{\phi}(\langle |\phi|\rangle)}-
      \frac{1}{e^{2}_{\phi}(\Lambda_{UV})}\right)\right] \, 
  \simeq \, \Lambda_{UV}\exp \left[ \frac{-24\pi^2}{e^{2}_{\phi}(\langle |\phi|\rangle)}
  \right]\,.
\end{equation}
We see that the vev $\langle |\phi|\rangle$ is generated at the scale
which is exponentially smaller than the UV cutoff scale $\Lambda_{UV}$
(in our case the Landau pole of $e^{2}_{\phi}$). Equation
\eqref{run2} is the consequence of the dimensional transmutation: the
dimensionality of the vev is carried by the UV-scale Landau pole,
while the exponential smallness of the ratio $\langle |\phi|\rangle /
\Lambda_{UV} \ll 1$ is guaranteed by the perturbativity of the
coupling constant $e^{2}_{\phi}$ in the vacuum {\it i.e.} at the scale
$\langle |\phi|\rangle$.  This addresses the naturalness problem.

\begin{figure}[t]
   \centering
   \includegraphics[width=9cm]{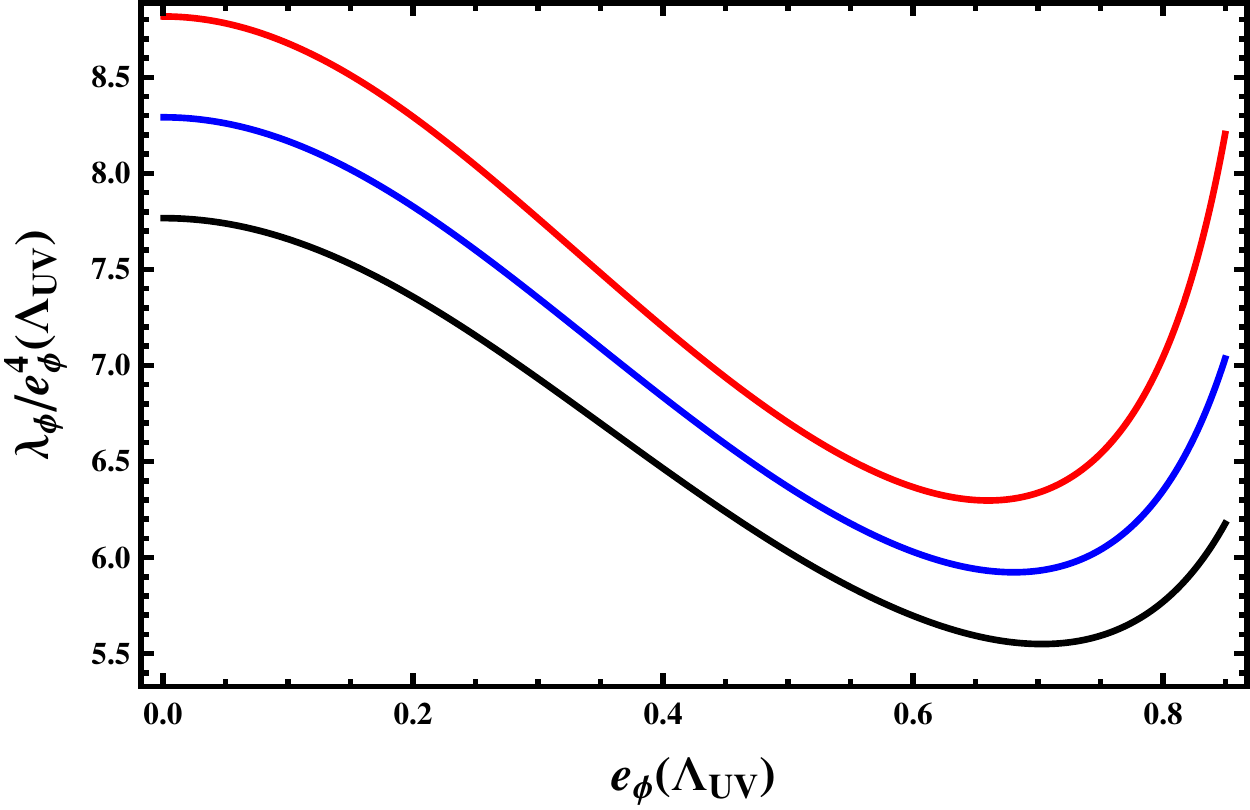} 
   \caption{\small Ratio of $\lambda_{\phi}/e^{4}_{\phi}$ at $\Lambda_{\rm
       UV}$ required to generate a hierarchy of $\Lambda_{\rm
       UV}/\langle\phi\rangle=10^{16},10^{15},10^{14}$ (from top to
     bottom) as a function of the gauge coupling $e_{\phi}$.}
   \label{fig:ratio}
\end{figure}

We would also like to quantify the exponential sensitivity of the vev
$\langle |\phi|\rangle$ to the {\it input} (or bare) values of the
coupling constants at the UV cutoff scale (we continue calling it
$\Lambda_{UV}$ even though here we don't think of it as a Landau
pole).  We proceed by solving the RG equation for the ratio of
coupling constants, which is obtained by combining \eqref{run1} with
the RG equation for $\lambda_{\phi}$,
\begin{equation}
  \label{run1lam}
  \frac{d \lambda_{\phi} }{dt} \,= \, \frac{1}{48 \pi^2} 
  \left(9 e^{4}_{\phi} - 3 e^{2}_{\phi} \lambda_{\phi}+
    \frac{5}{6} \lambda_{\phi}^2 \right)\,.
\end{equation}
For the ratio $x := {8\pi^2\lambda_{\phi}}/{(33 e^{4}_{\phi})}$ we find,
\begin{equation}
  \label{run1x}
  \frac{d x(t) }{dt} \,= \,
  \frac{6}{11}+x(t){\mathcal{O}}(e^{2}_{\phi}) + x(t)^2{\mathcal{O}}(e^{4}_{\phi})\,.
\end{equation}
Upon integration and setting the RG scale $M=\langle |\phi|\rangle$
and using the condition $x(\langle\phi\rangle)=1$ we obtain (we keep
only the first term on the {\it r.h.s.} of \eqref{run1x})
\begin{equation}
  \label{run2x}
  \langle |\phi|\rangle \,=\,\Lambda_{UV} \exp\left(\frac{11}{6}\right)
  \exp\left(-\frac{4\pi^2}{9}
    \frac{\lambda_{\phi}(\Lambda_{\rm UV})}{e^{4}_{\phi}(\Lambda_{\rm UV})}\right)\,.
\end{equation}
This expression shows the exponential sensitivity of the vev to the
values of the couplings at the UV cutoff.  As a result, the vev can
easily be made exponentially smaller than the UV cutoff (in agreement
with what we have already concluded from \eqref{run2}).  Qualitatively
the same behaviour holds beyond our simple approximation to the RG
equations.  This is shown in Fig.~\ref{fig:ratio}, where we show the
ratio of $\lambda/e^{4}_{\phi}$ at $\Lambda_{\rm UV}$ required to
generate a hierarchy of 14, 15 or 16 orders of magnitude between
$\Lambda_{\rm UV}$ and $\langle \phi\rangle$.

In summary, in a theory with no input mass scales, the
Coleman-Weinberg (CW) mechanism generates a symmetry-breaking vev and
the mass for the associated scalar from radiative corrections. These
scales are natural in the sense that they are automatically
exponentially suppressed compared to the UV scale at which we
initialise the theory.  Phenomenologically, however, this scenario has
a fatal flaw: if $\phi$ is the Higgs, then the Higgs mass turns out to
be too small. This is because the $\phi$ self-coupling is much smaller
than the gauge coupling $\lambda_{\phi} \ll e_{\phi}^2$. From
Eq.~\eqref{CWpotential} one can calculate the physical mass of the
Higgs remaining after spontaneous breaking of the gauge symmetry by
shifting the field $\phi = \langle \phi \rangle + \varphi$,
\begin{equation}
  \label{phimass}
  m^{2}_{\varphi}\,=\, \frac{3 e^{4}_{\phi}}{8\pi^2}\, \langle
  |\phi|^{2}\rangle \,.
\end{equation}
In terms of the mass $m^2_{X}=e^2_{\phi}\langle |\phi|^{2}\rangle$
of the vector boson we have
\begin{equation}
  \label{phimass2}
  m^{2}_{\varphi}\,=\, \frac{3 e^{2}_{\phi}}{8\pi^2}\,  m^{2}_{X} \, \ll \, m^{2}_{X}\,.
\end{equation}
This is in conflict with the observation that, in the Standard Model,
the Higgs is heavier than the corresponding vector bosons.

To resolve this problem we thus need to look beyond the
minimal Standard Model.  In this paper we consider a very compact
extension of the Standard Model where there is no longer a direct link
between the Higgs mass and the SM vector boson masses, and
consequentially, the Higgs can take its observed value $\sim
125$~GeV. At the same time, this formulation maintains the essential
feature that all mass scales are generated radiatively through
breaking of classical scale invariance via running couplings.

\medskip In the section~{\bf \ref{sec:two}} we outline the minimal
model we want to study: a scale-invariant Standard Model with an
additional CW ``hidden sector" and the Higgs portal-type coupling to
the SM.  In section~{\bf \ref{Phenomenology}} we analyse the
phenomenology of this model in the context of LHC and future
colliders, and low energy measurements.  We point out that with the
Higgs mass now being a known quantity, the minimal model has only two
remaining free parameters, and we show that the model is perfectly
viable.  The presently available Higgs data provide valuable
constraints on the parameter space, while future experimental data on
Higgs decays (as well as resonance searches) will further constrain
model parameters, and will ultimately provide discovery potential for
this model.

In a pre-LHC context this simple model has already been discussed
in~\cite{Hempfling:1996ht,Chang:2007ki} along with a variety of other
similar scalar field models in~\cite{Meissner:2006zh,Iso:2009ss}.
First model-building implications of a $\sim 125$~GeV Higgs have also
been looked at in~\cite{Iso:2012jn}.

\medskip

The use of the Coleman-Weinberg mechanism for BSM model building is
motivated by and based on the concept of classical scale
invariance. Although the scale invariance symmetry is anomalous, it
has been argued in~\cite{Bardeen:1995kv} that it may indeed be used as
a model-building guide and to motivate Coleman-Weinberg type
models~\cite{Meissner:2006zh}.  In section~{\bf \ref{sec:three}} we
provide another, renormalisation-group-inspired argument in favour of
this model-building strategy.

\medskip
Our conclusions are summarised in section~{\bf \ref{sec:five}}.

\section{Coleman-Weinberg with a Higgs portal}
\label{sec:two}

As we have seen in the previous section the main problem of the CW
scenario is that the mass of the Higgs boson is too small within the
SM.  The reason is that the mass of the Higgs is directly linked to
the mass of the gauge bosons and is 1-loop suppressed compared to
those.  A simple way to address this issue is to generate the mass
scale in a ``hidden sector'' and then transmit it to the SM, where it
directly acts as the scale $\mu^{2}_{\rm SM}$ of the pure SM.  This
breaks the direct CW relation between the SM gauge boson masses and
the mass of the SM Higgs boson.

A simple model to realise this is a Higgs-portal
model~\cite{Higgs.portal} with the CW toy model as a hidden
sector~\cite{Hempfling:1996ht,Chang:2007ki}. The classical potential
for scalar fields is,
\begin{equation}
  \label{potentialcoupled}
  V_{\rm cl}(H,\phi)\,=\, \frac{\lambda_{\rm H}}{2}(H^{\dagger} H)^{2}
  \,-\, \lambda_{\rm P}(H^{\dagger}H)|\phi|^{2}
  \,+\,\frac{\lambda_{\phi}}{4!}|\phi|^{4}\,.
\end{equation}
The first and the last terms are just the ordinary self-couplings for
the Higgs field and $\phi$ field, while the second term is the
Higgs-portal, coupling the SM Higgs field to the hidden sector field
$\phi$. For future convenience we chose the sign in front of this
Higgs-portal coupling to be negative.

To check the stability of this potential we complete the square in
\eqref{potentialcoupled}
  \begin{equation}
    \label{potentialcoupled2}
    V_{\rm cl}(H,\phi)\,=\, \frac{\lambda_{\rm H}}{2}\left(H^{\dagger} H
      \,-\, \frac{\lambda_{\rm P}}{\lambda_{\rm H}}|\phi|^{2}\right)^2
    \,+\, \frac{1}{24\lambda_{\rm H}}\left(\lambda_{\phi}\lambda_{\rm H}-12\lambda^{2}_{\rm P}\right)
    |\phi|^{4}\,.
  \end{equation}
The potential is then stable as long as
\begin{equation}
  \label{stab12}
  \lambda_{\phi}\lambda_{\rm H}>12\lambda^{2}_{\rm P}\,.
\end{equation}
When $\lambda_{\rm P} \to 0$ the two sectors decouple.

For non-vanishing $\lambda_{\rm P}$ the Higgs portal interaction can
generate the Higgs mass parameter of \eqref{VH} via
\begin{equation}
  \label{stability}
  \mu^{2}_{\rm SM} =- \lambda_{\rm P}\langle |\phi|^{2}\rangle\,.
\end{equation}

Importantly, in Eq.~\eqref{potentialcoupled} we have not allowed for
any mass terms. In other words we have a completely scale-free
potential even in presence of the Higgs portal coupling.  We now
proceed with employing the Coleman-Weinberg mechanism in the
Higgs-portal theory \eqref{potentialcoupled} where the complex scalar
$\phi$ is coupled as before to a U(1)$_{\rm hidden}$ gauge theory
(this forms the hidden sector), while the Higgs doublet $H$ has
standard interactions with the SU(2)$\times$U(1) gauge fields (as well
as matter fields) of the Standard Model.  At the origin in field
space, {\it i.e.} when all field vevs are zero, there are no scales
present in the classical scale-invariant theory.  We want to and can
preserve this feature in the quantum-corrected full effective
potential even after renormalisation by using\footnote{The term
  $\partial^{2} V(H,\phi)/\partial H^{\dagger} \partial
  \phi|_{H=\phi=0}$ vanishes by gauge invariance.}
\begin{equation}
  \label{rencond}
  \frac{\partial^{2} V(H,\phi)}{\partial H^{\dagger} \partial H}\bigg|_{H=\phi=0}=0\,,\quad
  \frac{\partial^{2} V(H,\phi)}{\partial \phi^{\dagger} \partial \phi}\bigg|_{H=\phi=0}=0\,.
\end{equation}
This is the same subtraction scheme as in the simple case
\eqref{eq:m0}, and as there, these conditions are automatic in
dimensional regularisation of any theory with classical scale
invariance.  In other regularisation schemes one cancels quadratic
divergencies between the bare masses and the counterterms. We
elaborate on this in more detail in next section.

The easiest way to visualise the emergence of electroweak symmetry
breaking in this theory is to consider a near decoupling limit.  If
$\lambda_{\rm P}\ll 1$ we can essentially view the process of symmetry
breaking independently in the two different sectors and we can view
electroweak symmetry breaking effectively as a two step process.

In the first step the CW mechanism generates a (large\footnote{Large
  compared to the electroweak scale of the standard model.}) vev
$\langle \phi\rangle$ in the hidden sector through dimensional
transmutation precisely as was outlined in the previous section.  In
the second step the vev $\langle \phi\rangle$ is transmitted to the
Standard Model via the Higgs portal, generating an effective mass
parameter for the Higgs
\begin{equation}
  -\mu^{2}_{\rm SM}\,=\,\lambda_{\rm P}\langle |\phi|^{2}\rangle
\end{equation}
Equation \eqref{musm} dictates that $\mu^{2}_{\rm SM}$ fixes the
electroweak scale, specifically,
\begin{equation}
  \label{musm2}
  -\mu^{2}_{\rm SM}\, =\, \frac{1}{2}\,m^{2}_{h}\,=
  \, \frac{1}{2}(125~{\rm GeV})^{2} \quad {\rm and} \quad
  - \mu^{2}_{\rm SM}\, =\, \frac{1}{2}\,\lambda_{\rm H} \, v^2
  \,\equiv\, \lambda_{\rm H} \, \langle |H|^{2}\rangle
\end{equation}
This implies that when $\lambda_{\rm P}\ll 1$ and also is much smaller
than other SM Higgs couplings, the electroweak scale is suppressed
compared to the hidden sector scale, as was anticipated,
\begin{equation}
  \label{musm3}
  \langle |\phi|^{2}\rangle \, =\, \frac{1}{\lambda_{\rm P}} \, 
  \frac{1}{2}(125~{\rm GeV})^{2}  \, =\, 
  \frac{\lambda_{\rm H}}{\lambda_{\rm P}} \, \langle |H|^{2}\rangle\,.
\end{equation}
The fact that the generated electroweak scale is much smaller than
$\langle \phi\rangle$ guarantees that any back reaction on the hidden
sector vev $\langle |\phi|^{2}\rangle$ is negligible.

Let us now verify that the dimensional transmutation phenomenon
continues to work in our more complicated theory and all the required
vevs are natural.  To see this we start from the Higgs-portal
effective potential
\begin{equation}
  \label{Veff2hp}
  V(\phi,H)\,=\, \frac{\lambda_{\phi}}{4!}|\phi|^{4}\,+\,
  \frac{3 e_{\phi}^{4}}{64\pi^2}|\phi|^{4}
  \left[\log\left(\frac{|\phi|^{2}}{\langle |\phi|^{2}\rangle}\right)-\frac{25}{6}\right]
  \,-\, \lambda_{\rm P}(H^{\dagger}H)|\phi|^{2}\,+\,
  \frac{\lambda_{\rm H}}{2}(H^{\dagger} H)^{2}\,.
\end{equation}
Here we are keeping 1-loop corrections arising from interactions of
$\phi$ with the U(1) gauge bosons in the hidden sector, but neglecting
radiative corrections from the Standard Model sector. The latter would
produce only subleading corrections to the vevs.  The
$\phi$-minimisation condition\footnote{Minimisation with respect to
  $H$ does not give anything new beyond the known SM condition
  \eqref{musm}.}  for this effective potential is ({\it cf.}
\eqref{Veffprime} and \eqref{musm3})
\begin{equation}
  \label{Veffprime2}
  \partial_{\phi} V\,=\, \frac{1}{6}\left( \lambda_{\phi}-\frac{33}{8\pi^2}e^{4}_{\phi}\right)
  {\langle \phi\rangle}^3   - 2\lambda_{\rm P}{\langle |H|^2 \rangle}{\langle \phi\rangle}\, =\,
  \frac{1}{6}\left( \lambda_{\phi}-\frac{33}{8\pi^2}e^{4}_{\phi}-12\frac{\lambda_{\rm P}^2}{\lambda_{\rm H}}\right)
  {\langle \phi\rangle}^3  \, =\, 0
\end{equation}
We thus conclude that the dimensional transmutation continues to work
and the and the vev $\langle \phi\rangle$ is determined by the
condition on the four couplings renormalised at the scale of the vev
\begin{equation}
  \label{eq:rad4}
  \lambda_{\phi}(\langle|\phi|\rangle)-\frac{33}{8\pi^2}e^{4}_{\phi}(\langle |\phi|\rangle)
  -12\frac{\lambda_{\rm P}^2 (\langle |\phi|\rangle)}{\lambda_{\rm H}(\langle |\phi|\rangle)} \,=0 \,.
\end{equation}
For small $\lambda_{\rm P}$, this is a small deformation of the
original condition \eqref{eq:rad4}. In the near-decoupling case of
$\lambda_{\rm P} \ll 1$ we are interested here, the modifications are
negligible.  But even in a more general case, there are no
obstructions for the Coleman-Weinberg mechanism to work.  

The two vevs, $\langle \phi\rangle$ and $v$ are generated {\it
  naturally} through dimensional transmutation in our framework
similarly to \eqref{run2},
\begin{equation}
  \label{run3}
  \sqrt{\frac{\lambda_{\rm H}}{\lambda_{\rm P}}} \, \langle H \rangle \,=\,
  \langle \phi \rangle  \, \simeq \, \Lambda_{UV}\exp 
  \left[ \frac{-24\pi^2}{e^{2}_{\phi}(\langle |\phi|\rangle)}
  \right] \, \ll \Lambda_{UV}\,.
\end{equation}
Since massive vector bosons of the Standard Model play no role in
stabilising the minimum of the Coleman-Weinberg potential in our Higgs
portal model, there is no condition linking the SM gauge and the Higgs
couplings. As a result the vector boson masses and the Higgs boson
mass are independent and can take their observed SM values.

\section{Arguments in favour of vanishing mass terms at the origin of the potential}
\label{sec:three}

The exponential sensitivity of the Higgs vacuum expectation value to
the boundary values of the couplings, and the natural generation of
the hierarchy between the EWSB scale and cut-off scale in
\eqref{run3}, crucially depend on the choice of massless
renormalisation conditions Eq.~\eqref{rencond} at the origin of the
field space. In this section we want to give arguments in favour of
this choice.

A suitable symmetry to forbid mass terms for scalars is scale
invariance.  Indeed, in absence of scale invariance the classical
potential Eq.~\eqref{potentialcoupled} would allow for two additional mass
terms,
\begin{equation}
  \Delta V_{\rm no\,\, scale\,\, invariance}=m^{2}_{\rm H}H^{\dagger}H+m^{2}|\phi|^{2}\,.
\end{equation}

In the class of theories we consider, scale invariance is a classical
symmetry which is broken by quantum corrections, specifically by the
logarithmic running of the couplings.  One might therefore query if it
is allowed to set these mass terms to zero in full quantum theory, as
we have done in Eq.~\eqref{rencond}.
In~\cite{Bardeen:1995kv,Meissner:2006zh} this question has been
answered favourably based on the special role played by dimensional
regularisation and considering the anomaly in the trace of the
energy-momentum tensor. Here, we provide additional perspective and
support based on the renormalisation group, and also address the
question of scheme dependence.

First we want to check if our requirement that the mass terms
vanish~\eqref{rencond}, is affected by a change in the renormalisation
scale. To do this we can look at the appropriate renormalisation group
equations for the mass terms.  In \emph{dimensional regularisation}
they have the form,
\begin{equation}
  \label{massrg}
  \partial_{t}\left(\frac{m^{2}_{i}}{M^2}\right)\equiv
  \partial_{t} \epsilon_{i}=(-2+\eta_{i})\epsilon_{i}\,,
\end{equation}
with $i=H,\phi$ and $\eta_{i}$ the anomalous dimension of the Higgs
and $\phi$ field respectively, and $t=\log M$ as before.

We can clearly see that $\epsilon_{i}=0$ is a fixed point of the RG
evolution and, once enforced at one scale, it holds at all scales. In
this sense -- within dimensional regularisation -- our renormalisation
conditions Eq.~\eqref{rencond} are self-consistent and contain no
fine-tuning.  They correspond to an enhanced unbroken symmetry for
these couplings.

In the argument above we made use of a specific regularisation scheme:
dimensional regularisation. In other regularisation
schemes\footnote{Most other schemes introduce a new mass scale which
  explicitly breaks scale-invariance.} the (one-loop) RG equations
have a different form,
\begin{equation}
\label{fxdpt}
  \partial_{t} \epsilon_{i}=(-2+\eta_{i})\epsilon_{i} + 
  c_{i, e}e^{2}_{\phi}+c_{i, \lambda_{\rm P}}\lambda_{\rm{p}}+c_{i, \lambda_{\phi}}\lambda_{\phi}
  \qquad {\hbox{not dimensional regularisation}}
\end{equation}
with constants $c_{i}$ that depend on the regularisation scheme.

The terms $\sim c_{i}$ destroy the fixed point at $\epsilon_{i}=0$.
Instead we now have a partial\footnote{I.e. it is a fixed point when
  we neglect the running of $e_{\phi}$, $\lambda_{\rm P}$ and
  $\lambda_{\phi}$.} fixed point at,
\begin{equation}
\epsilon_{i, \rm partial} = 
  \frac{c_{i, e}e^{2}_{\phi}+c_{i, \lambda_{\rm P}}\lambda_{\rm{P}}+c_{i, \lambda_{\phi}}\lambda_{\phi}}{(2-\eta_{i})}.
\end{equation}

Neglecting the evolution of $e_{\phi}$, $\lambda_{\rm P}$ and $\lambda_{\phi}$ we can now write the
RG equation for $\epsilon$ as
\begin{equation}
\partial_{t}(\epsilon_{i}-\epsilon_{i, \rm partial})=(-2+\eta_{i})(\epsilon_{i}-\epsilon_{i,\rm partial}).
\end{equation}
This equation has the simple solution,
\begin{equation}
  (\epsilon_{i}-\epsilon_{i, \rm partial})(t)=(\epsilon_{i}-\epsilon_{i, \rm partial})(t_{0})\exp[(-2+\eta_{i})(t-t_{0})]=(\epsilon_{i}-\epsilon_{i, \rm partial})(t_{0})\left(\frac{\Lambda}{M}\right)^{2-\eta_{i}}
\end{equation}
where the staring point for the trajectory is $t_{0}=\log(\Lambda)$,
so that the combination $t-t_0$ appearing above, is $\log(M/\Lambda)$.
We now let the trajectory run from the high scale $t_0$  to a low value of $M$.
At weak coupling we can neglect the anomalous dimensions. Using an initial value $\epsilon_{i}(t_{0})=0$ corresponding to a vanishing
mass at the scale $\Lambda$ we recover the usual
quadratic divergencies,
\begin{equation}
  \label{quad}
  m^{2}_{i}(\Lambda)-m^{2}_{i}(M\sim 0)\approx \frac{1}{2}(c_{e}e^{2}_{\phi}+
  c_{\lambda_{\rm P}}\lambda_{\rm P}+c_{\lambda_{\phi}}\lambda_{\phi})\Lambda^{2}\,.
\end{equation}

In all regularisation schemes with non-vanishing $c_{i}$, scale
invariance is broken more strongly than in dimensional regularisation.
We can therefore turn the argument around and argue that dimensional
regularisation, having no quadratic divergencies, is the scheme which
exhibits the smallest breaking of scale invariance.  If we now insist
that scale invariance is broken minimally by quantum corrections we
are automatically led to dimensional regularisation and therefore
Eq.~\eqref{massrg} and consequently to our renormalisation conditions
Eq.~\eqref{rencond}.

In absence of additional mass scales in the theory we are free to make
this choice. Indeed one can argue that this is a preferred choice
since in this case the only scale invariance breaking effect is the
logarithmic running of the dimensionless couplings, which is
independent of the regularisation scheme. All scheme-dependent (and
therefore unphysical) effects are set to zero.  \bigskip

Let us take a step back from our concrete model and take a look at the
more general situation. From a renormalisation group point of view,
consistent theories are those that have a UV fixed point in the space
of dimensionless coupling constants (all coupling constants of higher
dimensional operators can be made dimensionless by scaling with an
appropriate power of the RG scale, therefore this space is very
infinite dimensional). In order for a theory to be predictive we need
to be able to describe it by a finite number of parameters. For the UV
fixed point this means the following: the space of all RG trajectories
ending in the fixed point as the RG scale is taken to infinity is
finite dimensional. The number of these dimensions is the number of
free parameters. In the usual language these are the \emph{relevant}
parameters.  Exciting a combination of coupling constants that is not
in this subspace leads to an RG trajectory that (per definition) does
not end in the fixed point as we go into the UV and the theory has
incurable divergencies. The fixed point of a theory defined in this
manner can be in the perturbative region where all coupling constants
are small, but it can also be in a non-perturbative regime. In the
latter case we have a non-perturbatively renormalisable theory in the
sense of Weinberg's scenario of ``asymptotic safety''~\cite{Weinberg}.

To give a concrete example, consider QCD with one flavour of massive fermions. This
theory has two relevant parameters that can be chosen to be
non-vanishing, the gauge coupling constant $g$ and the mass $m$
divided by the RG scale $M$, $\epsilon=m/M$.  The RG equations are
\begin{equation}
  \label{RGexample}
  \partial_{t} g=0\times g-\frac{31}{3}\frac{g^{3}}{16\pi^2},\qquad
  \partial_{t}\epsilon=(-1+{\mathcal{O}}(g^{2}))\epsilon\,.
\end{equation}
One can easily check that starting from any (sufficiently small) value
of $g$ and $\epsilon$, we end up in the UV fixed point $g=\epsilon=0$.
However, for any ``non-renormalisable'' operator such as, {\it e.g.},
$a(G^{\mu\nu}G_{\mu\nu})^{2}$ with dimensionless coupling constant,
$\xi=a \cdot M^{4}$ 
we have,
\begin{equation}
  \label{RGexample2}
  \partial_{t}\xi=(+4+{\mathcal{O}}(g^{2}))\xi\,,
\end{equation}
which for any starting point with non-vanishing $\xi$ (but still
close enough to the fixed point) has a trajectory that rapidly moves
away from the fixed point. The same argument holds for any other
higher dimensional operator.

The terms on the right hand sides of Eqs.~\eqref{RGexample},
\eqref{RGexample2} that are linear in the couplings whose change is
described on the left hand sides, describe the approach to, or running
away from the fixed point in the UV\footnote{More precisely the
  $d_{X}$ are the eigenvalues of the stability matrix of the system of
  RG equations. Close to the perturbative fixed points, they are given
  by minus the naive dimension of the coupling plus its anomalous
  dimension.},
\begin{equation}
  \partial_{t} X=d_{X} \times X \quad \Rightarrow \quad X(M)=X(M_{0})\left(\frac{M}{M_{0}}\right)^{d_{X}}\,.
\end{equation}
Clearly, those directions with negative $d_{X}$, approach the fixed
point $X \to 0$ as the RG-scale $M$ goes to infinity; while those with
positive $d_{X}$, diverge.  The case with $d_{X}=0$ leads to the usual
marginal behaviour with logarithmic running towards (or away from) the
fixed point.

Going in the opposite direction towards smaller $M$, the operators with
$d_{X}<0$ are exactly those that quickly obtain very large values.
This is where the hierarchy problem lies. Choosing ``natural''
${\mathcal{O}}(1)$ initial values for those operators at some UV
scale, we get enormous values at a smaller scale. Vice versa, to get
a value ${\mathcal{O}}(1)$ at some small scale requires us to finely
tune the initial value at the high scale to be extremely small.

Importantly the $d_{X}$ are the critical exponents of the theory
which are thought to be {\emph{scheme-independent}}.

Our proposal is now as follows. Let us restrict our theory to live on
a subspace of all trajectories which end in the fixed point. This
subspace is defined by only exciting the marginal $d_{X}=0$
trajectories.  Then {\emph{all scales are generated via dimensional
    transmutation from the logarithmic running of the coupling
    constants.}}  This is not a fine-tuning because we require the
$d_{X}\neq 0$ operators to be {\emph{exactly zero}}, {\it i.e.} we are
living exactly on this well-defined subspace\footnote{While the
  precise shape of this subspace is scheme dependent, its existence
  and dimensionality is not.}.

Our concrete example now shows that we can choose the initial value of
the Higgs mass operator at the high scale to be vanishing while still
getting a phenomenologically viable non-vanishing vacuum expectation
value (and physical Higgs mass) by dimensional transmutation from the
marginal operators which exhibit only logarithmic
running.\footnote{There is a beauty defect in our theory in that the
  marginal couplings are actually marginally irrelevant, but one can
  hope to cure this by a suitable embedding in a more complete
  theory.}
The trajectories in the subspace requiring all $d_{X}=0$, are
exactly those that correspond to classical scale invariance, broken
only by the logarithmic running induced by quantum corrections.

\bigskip

Alternatively one can consider theoretical setups like Supersymmetry
(SUSY) where the quadratic divergences are absent\footnote{One could
  be even more ambitious and ask that the theory is finite but this
  does not change our argument.}. Such a theory has an additional
scale (SUSY-breaking scale) above which the quadratic divergencies are
canceled. This scale is physical in the sense that the dynamics of the
theory above this scale is qualitatively different from the behaviour
in the IR. This is for example due to the appearance of new degrees of
freedom at higher energies. In such a situation we get \emph{finite}
threshold corrections from these additional degrees of freedom in any
regularisation scheme.  Simulating the quadratic divergencies of
ordinary theories, these threshold corrections typically also scale
quadratically with the scale of new physics. In the case of SUSY,
above the SUSY-breaking scale quadratic divergences are canceled
between bosons and fermions, leaving threshold corrections to the
Higgs mass. These corrections are proportional to the mass squares of
the SUSY partners of the SM particles and therefore quadratically
sensitive to the scale at which SUSY is broken.

Even in this type of setup, our Coleman-Weinberg scenario is a helpful
step to bridge the (possibly large) gap to this scale of new physics
without generating a big fine-tuning. In the more complete theory we
then only need to ensure that the sum total of all the finite
threshold corrections vanishes. To us this seems a more achievable
goal then getting a small (compared to the scale of new physics) but
non-vanishing sum of threshold corrections.


\section{Phenomenology}
\label{Phenomenology}
Let us investigate in this section the phenomenological viability as
well as possible signatures of the proposed model.

In the hidden sector we have two additional fields $\phi$ and the
extra U(1)$_{\rm hidden}$ gauge field $X^{\mu}$.  After $\phi$ acquires a
non-vanishing vev the gauge field becomes massive with a mass
\begin{equation}
m_{X}=e_{\phi}\langle\phi\rangle\,.
\end{equation}
In principle this extra U(1)$_{\rm hidden}$ gauge boson can
kinetically mix~\cite{holdom} with the hypercharge U(1), allowing for
a rich phenomenology which can also be tested at the
LHC~\cite{Frandsen:2012rk}. Here we will not consider such a mixing
and instead focus only on those interactions that must be present in
order to ensure a working electroweak symmetry breaking, which will
also modify the Higgs phenomenology.

In absence of kinetic mixing the dominant interaction between the
hidden sector and the SM is via the Higgs portal coupling
$\lambda_{\rm P}$.  The lowest order effect arises from the mixing
between the SM Higgs $H^T(x)=\frac{1}{\sqrt{2}}(0, v+h(x))$ and the
hidden Higgs $\phi = \langle \phi \rangle + \varphi$.  The two
scalars, $h$ and $\varphi$, mix via the mass matrix,
\begin{equation}
\label{massmixing}
  m^{2}=\left(
\begin{array}{cc}
  m^{2}_{h}+\Delta m^{2}_{h,\rm SM} & -\kappa \,m^{2}_{h} \\
  -\kappa\, m^{2}_{h}  &  m^{2}_{\varphi} +\kappa^{2} m^{2}_{h}
\end{array}\right)\,,
\end{equation}
with the mixing parameter,
\begin{equation}
\kappa=\sqrt{\frac{2\lambda_{\rm P}}{\lambda_{H}}}\,,
\end{equation}
and the masses
\begin{equation}
  m^{2}_{h}=\lambda_{\rm H} v^{2},\qquad 
  m^{2}_{\varphi}=\frac{3e^{4}_{\phi}}{8\pi^2}\langle\phi\rangle^{2}
  =\frac{3 e^{2}_{\phi}}{8\pi^2}m^{2}_{X}\,.
\end{equation}
These are the same as we had in the decoupled case ($\lambda_{\rm P}
=0=\kappa$) for the Higgs mass and the CW scalar $\varphi$ mass.

In Eq.~\eqref{massmixing} we have also included one-loop corrections
to the SM Higgs mass,
\begin{equation}
  \Delta m^{2}_{h,\rm{SM}}=\frac{1}{16\pi^2}\frac{1}{v^2}\left(6m^{4}_{W}+3m^{4}_{Z}+m^{2}_{h}-24m^{4}_{t}\right)
  \approx -2200\, {\rm GeV}^{2}
\end{equation}
Numerically, these corrections are dominated by the
top-quark loop and are therefore {\rm negative}.  While the resulting
contributions are small for the nearly decoupled limit and at large
$m^{2}_{\varphi}$, they lead to interesting effects for the case of
small $m^{2}_{\varphi}$ and moderate Higgs portal coupling.

Depending on the CW mass scale induced in
the hidden sector, the model predicts new resonant structures in
di-Higgs analysis or a hidden Higgs decay phenomenology as main
modifications of the electroweak sector compared to the SM.

This matrix can be easily diagonalised with a rotation,
\begin{equation}
  \label{eq:rot}
  \left(
    \begin{array}{c}
      h_{1}\\
      h_{2}
    \end{array}
  \right)=
  \left(\begin{array}{cc}
      \cos\vartheta & \sin\vartheta\\
      -\sin\vartheta & \cos\vartheta
    \end{array}
  \right)
  \left(\begin{array}{c}
      h\\
      \varphi
    \end{array}
  \right),\qquad {\rm with}\quad \vartheta
  \approx\kappa\frac{m^{2}_{h}}{m^{2}_{\varphi}-m^{2}_{h}-\Delta m^{2}_{h,\rm SM}}\ll1 \,,
\end{equation}
where the right hand side in the definition of gives $\vartheta$ in
the case of small mixing.

Up to order $\vartheta^2$ ({\it i.e.} to leading order in $\lambda_{\rm P}$)
the masses of the two eigenstates are simply
\begin{equation}
\label{masslowest}
m^{2}_{h_{1}}=(m^{2}_{h}+\Delta m^{2}_{h,\rm SM}) (1+{\cal O} (\vartheta^2)) ,
\qquad m^{2}_{h_{2}}=m^{2}_{\varphi}(1 +{\cal O} (\vartheta^2))\,.
\end{equation}

Fixing the (dominantly SM like) state $h_{1}$ to have a mass of $\sim
125~{\rm GeV}$ we can now look at possible constraints on the only two
remaining parameters: the mixing angle $\vartheta$, and the mass of
the second eigenstate. Due to the rotation \eqref{eq:rot} the model
will show the character traits of a Higgs portal
model~\cite{Higgs.portal,portrefs,hidden1,hidden2,singletn}, however
with restrictions on the parameters that follow from
transmitting EWSB to the visible sector as laid out in the previous
sections.

Let us enumerate the parameters of our model in the small
$\lambda_{\rm P}$ regime we are working in.  The SM Higgs
self-coupling is fixed by the ratio of known electroweak scales, while
the other self-coupling, $\lambda_{\phi}$, is determined from the CW
dimensional transmutation condition:
\begin{equation}
\label{self-cs}
\lambda_{H}\,=\, \left(\frac{m_h}{v}\right)^2 \,\approx\, \frac{1}{4}
 ,\qquad \lambda_{\phi}\,=\, \frac{33}{8\pi^2} \, e^4_{\phi} \,.
\end{equation}
There are two undetermined parameters in our model which one can take
to be the hidden sector gauge coupling, $e^2_{\phi}$, and the (small)
portal coupling $\lambda_{\rm P}$. In this case, the two mass scales
associated with the hidden scalar are fixed,
\begin{equation}
  \label{hid-scs}
  \langle |\phi|^2 \rangle \,=\, \frac{1}{2 \lambda_{\rm P}}\, m_h^2
  ,\qquad
  m_{\varphi}^2 \,=\,\frac{3 e^4_{\phi}}{8\pi^2}\,\langle |\phi|^2
  \rangle \,=\,
  \frac{3 e^4_{\phi}}{16\pi^2}\,\frac{1}{\lambda_{\rm P}}\, m_h^2\,,
\end{equation}
and the hidden sector vector mass is given by
\begin{equation}
  m^{2}_{X}
  \,=\,\frac{8\pi^2}{3e^{2}_{\phi}}\,m^{2}_{\varphi}
  =\frac{e^{2}_{\phi}}{2\lambda_{\rm P}}m^{2}_{h}\,.
\end{equation}

Alternatively, the two free parameters can be chosen to be the mass of
the hidden Higgs, $m_{\varphi}$, and the Higgs portal coupling
$\lambda_{\rm P}$.  In this case, gauge coupling (and the mixing
$\kappa$) is determined via
\begin{equation}
  \label{hid-sclp}
  e^{2}_{\phi}=\frac{4\pi}{\sqrt{3}}\sqrt{\lambda_{\rm
      P}}\frac{m_{\varphi}}{m_{h}}\,,\quad{\rm and}\quad 
  \kappa=\sqrt{\frac{2\lambda_{\rm P}}{\lambda_{H}}}\,.
\end{equation}

In analogy to the Standard Model sector one may expect $e_{\phi}$ be
of order $0.1-1$ but it could also be much smaller ({\it cf.} the
hyperweak interactions in~\cite{Burgess:2008ri}). In the latter case
one would, however, also need to explain an incredibly small
$\lambda_{\phi}$. More importantly, for small $m^{2}_{\varphi}$ it is
crucial to take into account higher order corrections to
Eq.~\eqref{masslowest},
\begin{equation}
  \label{massmin1}
  m^{2}_{h_{2}}=m^{2}_{\varphi}(1+{\mathcal{O}}(\vartheta^{2}))
  +\vartheta^2 \frac{\Delta m^{2}_{h,\rm SM}}{m^{2}_{h}} (m^{2}_{h}+\Delta m^{2}_{h,\rm SM}).
\end{equation}
As the SM model correction $\Delta m^{2}_{h,\rm SM}$ is negative (and
quite sizeable) this enforces a minimal value for $m^{2}_{\varphi}$
from the stability requirement $m^{2}_{h_{2}}>0$,
\begin{equation}
  \label{massmin}
  m^{2}_{\varphi}\geq m^{2}_{\varphi,\rm min}= \frac{2\lambda_{\rm P}}{\lambda_{H}}\left(\frac{|\Delta m^{2}_{h,\rm SM}|}{m^{2}_{h}+\Delta m^{2}_{h,\rm SM}}\right)m^{2}_{h}.
\end{equation}

The minimal mass for $m^{2}_{\varphi}$ can also be translated into a
minimal value for
\begin{equation}
  e^{2}_{\phi}\geq \frac{2\lambda_{\rm P}}{\lambda_{H}}
  \sqrt{\lambda_{H}\frac{8\pi^2}{3}\frac{|\Delta m^{2}_{h,\rm SM}|}{m^{2}_{h}+\Delta m^{2}_{h}}}.
\end{equation}
and, more importantly, into the lower bound for the mass of the
U(1)$_{\rm hidden}$ gauge boson,
\begin{equation}
  m^{2}_{X}
  \geq m^{2}_{h}\sqrt{\frac{1}{\lambda_{H}}\frac{8\pi^2}{3}\frac{|\Delta m^{2}_{h,\rm SM}|}{m^{2}_{h}+\Delta m^{2}_{h}}}\approx 250\,{\rm GeV}.  
\end{equation}

For the physical mass $m^{2}_{h_{2}}$ Eqs.~\eqref{massmin1} and
\eqref{massmin} also entail that values much below $m^{2}_{\varphi,\rm
  min}$ require some amount of fine-tuning as it involves a
cancellation between $m^{2}_{\varphi}$ and the SM correction.
Moreover, for very small hidden Higgs masses the mixing is very
strongly constrained from fifth force
measurements~\cite{Ahlers:2008qc}.  In the following we will therefore
mostly concentrate on the case of moderate $e_{\phi}\sim 0.1\dots 1$
and hidden Higgses with masses $m_{\phi}\gtrsim {\rm MeV}$.

\begin{figure}
  \begin{center}
    \includegraphics[width=0.6\textwidth]{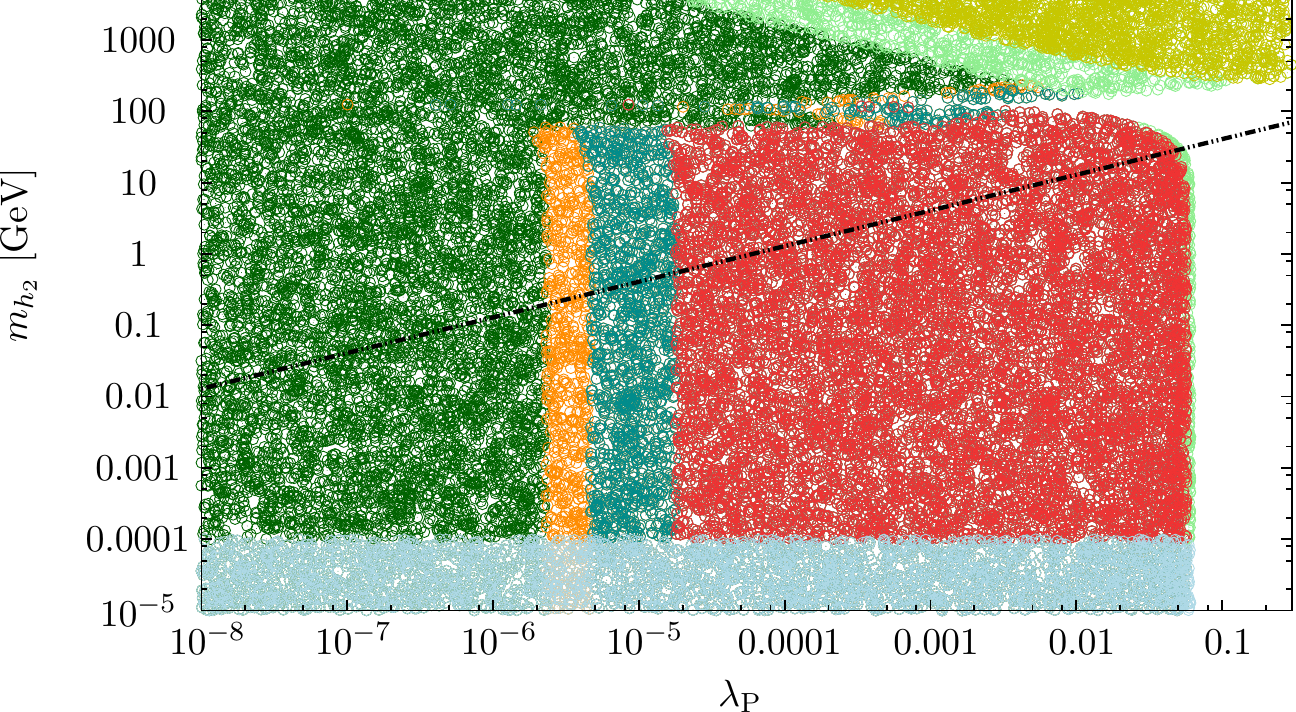}\hfill
    \caption{\small \label{constraints} Scatter plot of the model
      described in Sec.~\ref{Phenomenology} for $10^5$ randomly
      generated parameter choices in the $(\lambda_{\rm{P}},m_{h_2})$
      plane. Points below the black dash-dotted line require some
      fine-tuning according to Eqs.~\eqref{massmin1},
      \eqref{massmin}. The region excluded by current LHC measurements
      is shown in red.  The cyan region can be probed by LHC with high
      luminosity and the orange region shows a projection for a
      combination of a high luminosity LHC with a linear
      collider. Light blue indicates constraints from stellar
      evolution.  The constraints on the parameter space for a Landau
      pole separation of 4, and 16 orders of magnitude are included in
      yellow and light green, respectively. The remaining allowed
      parameter points are depicted in green.}
  \end{center}
\end{figure}

A first constraint can be imposed from theoretical reasoning. From
Eq.~\eqref{eq:rad4} we can see that $\lambda_{\phi}$ grows as
$e_{\phi}$ and/or $\lambda_{\rm P}$ are increased raising the
possibility of a nearby Landau Pole. Requiring that there is no Landau
pole in $\lambda_{\phi}$ for at least a few orders of magnitude puts
already fairly strict limits on both $\lambda_{\phi}$ and
$\lambda_{\rm P}$. Neglecting the $\lambda^{2}_{\rm P}$ contribution
to the running of $\lambda_{\phi}$ the solutions to the RG equations
are given in~\cite{Coleman:1973jx}. In Fig.~\ref{constraints} we show
the constraints arising from a hierarchy of 4 and 16 orders of
magnitude between $\langle\phi\rangle$ and the Landau pole yellow and light green, respectively. 
We can see that
this automatically restricts us to fairly small
$\lambda_{\text{P}}$. The used approximation is conservative in the
sense that the $\lambda^{2}_{\rm P}$ contribution to the running of
$\lambda_{\phi}$ is positive speeding up the approach to the Landau
pole. On the other hand for small $\lambda_{\rm P}$ the neglected term
quickly becomes very small.

If the mass $m_{h_1}>2m_{h_2}$ the ordinary Higgs can decay into two
hidden Higgses.  In leading order in the mixing angle this decay
occurs via the term
\begin{equation}
  {\mathcal{L}}\supset - \lambda_{\rm P} v h\varphi^{2}\,.
\end{equation}
and the SM-like Higgs trilinear interaction. The rotation to the
physical mass eigenstates $h_1,h_2$ complicates the formulea, i.e. the
trilinear couplings get more involved (see
e.g. Refs.~\cite{portrefs,Dolan:2012ac} for detailed discussion). We
fully include these nonlinear effects in our later scan but only
sketch the line of thought in the following which is valid for small
mixing, i.e $h_2\sim \varphi$.  Hence, the dominant part of the
corresponding partial decay width is
\begin{equation}
  \label{eq:decay}
  \Gamma_{h_{1}\to h_{2}h_{2}}
  =\frac{{4}\lambda^{2}_{\rm P}v^2}{16\pi}
  \frac{[m^{2}_{h_{1}}-4m^{2}_{h_{2}}]^{1/2}}{m^{2}_{h_{1}}}\,,
\end{equation}
and needs to be taken into account for the Higgs modified branching
ratios (BRs). A similar equation holds for $m_{h_{2}}>2m_{h_{1}}$ with
$v\rightarrow \sqrt{2}\left\langle|\phi|\right\rangle$ and $m_{h_{1}}
\leftrightarrow m_{h_{2}}$. In our simple setup there are no light hidden
sector particles into which the hidden Higgs can decay. The $h_2$
therefore decays back into SM particles via the mixing with the Higgs
and its couplings to light particles. The branching ratios are the
same as for the SM Higgs with mass $m^{2}_{h_{2}}$, but the width, as
well as the production cross sections from visible matter, are reduced
by a factor $\sin^{2}\vartheta$,
\begin{align}
  \Gamma_{h_{2}\to XX^c}=&\sin^{2}\vartheta\,\Gamma^{\rm SM}_{h\to
    XX^c}(m_{h}=m_{h_{2}})\,,\\
  \sigma(XY\to h_{2} )=& \sin^{2}\vartheta \,\sigma^{\rm SM}_{XY\to
    h}(m_{h}=m_{h_{2}})\,.
\end{align}
Note, that already the SM Higgs decay width is quite small,
$\Gamma_{\rm SM}(m_{h}\simeq 125~{\rm GeV}) \simeq 4~{\rm MeV}$
\cite{hwidth} and decreases more or less linearly (until the bottom
threshold is crossed) with the mass. Combining this with a small
mixing angle, $h_2$ becomes an extremely narrow resonance. Indeed
for very small values of $\vartheta$ we may even have displaced
vertices or can use adapted trigger strategies~\cite{trigger} to
constrain such a scenario at the LHC (signatures of this type have
been described in~\cite{Englert:2011yb}).

In Fig.~\ref{constraints} we show the results of a parameter scan
projected on the $(\lambda_{\rm{P}},m_{h_2})$ plane (we identify
$m_{h_1}\simeq 125~\text{GeV}$). We include constraints from current
LHC searches for the mass range $m_{h_2}\gtrsim 114~{\rm{GeV}}$, which
can be as low as $\sigma\times\text{BR}\simeq 0.1$~\cite{newbounds}
and the LEP constraints for $m_{h_2}\lesssim 114~{\rm{GeV}}$,
precision constraints from the $S,T,U$ parameters~\cite{peskin} as
well as tree-level unitarity constraints are imposed. The currently
allowed coupling span of the Higgs measurements is $\sigma\times
{\text{BR}}/\left[\sigma\times {\text{BR}}\right]_{\text{SM}}\gtrsim
0.7$ at 1 sigma~\cite{plehnrauchlhc}, which is the combined result of
the discovery channels $h\to WW,ZZ,\gamma\gamma$. In our model we
always have $\sigma\times {\text{BR}}/\left[\sigma\times
  {\text{BR}}\right]_{\text{SM}}< 1$ due to mixing; a statistically
significant measurement of the enhancement in the $h\to \gamma\gamma$
would therefore be at odds with the most straightforward
implementation of EWSB as described in the preceding sections.

A significant decay of the Higgs candidate into fermions is yet to be
measured. Current constraints on $h\to b\bar b$ (with SM branching
ratio $\simeq 60\%$) follow from biasing the coupling fit with the SM
assumption of a total SM-like Higgs decay width. The observed rates,
at the current precision can be understood as a limit on the total
Higgs width itself. Given that we have a potentially large coupling to
a new decay channel at a large available phase space
Eq.~\eqref{eq:decay} an upper limit on the total Higgs width
constrains the model. Recent analyses suggest
$\Gamma_h/4~\text{MeV}\lesssim 1.3$~\cite{Dobrescu:2012td} and we
include this bound to our scan. We also display the improvement of the
ruled-out region due to the combination of a high-luminosity LHC run
in combination with a linear collider on the basis of the most recent
coupling fits of Ref.~\cite{Klute:2013cx}. Note that, other than at a
hadron collider, the total Higgs width can be measured by correlating
Higgs production in weak boson fusion $e^+e^-\to \nu\bar\nu h$ and the
decay $h\to WW$ at the $\Gamma_h/4~\text{MeV}\lesssim 10\%$
level~\cite{ilch,hidden1}.

From Fig.~\ref{constraints} we see that there is a large parameter
region of the model allowed by current measurements (note that the
allowed region, of course further extends to smaller
$\lambda_{\text{P}}$ and also to larger masses).  The model can be
efficiently constrained by measuring the Higgs candidates cross
section and decay width as precisely as possible, which can be done
extraordinarily well at a precision collider instrument such as a
future linear collider. The small funnel region at around
$m_{h_2}\approx m_h = 125~\text{GeV}$ follows from relaxed bounds and
kinematic suppersion in the vicinity of the Higgs
candiadate. $m_{h_2}$ within this range is then unconstrained more or
less irrespective of the precise value of $\lambda_{\rm {P}}$.

Since, $\lambda_{\rm{P}}$ is small by consistency and RG arguments, we
face small mixing with the hidden sector which effectively yields a
phenomenologically decoupled Higgs partner in the single Higgs
channels of \cite{ATLAS:2012gk,CMS:2012gu} when background
uncertainties are taken into account. When the mixing is rather larger
sensitivity in SM-like Higgs searches can provide powerful means to
constrain the model for heavy $m_{h_2}$. Given the small width,
standard analyses can be straightforwardly extended beyond the current
upper limit of $m_{h_1}\leq 1~{\text{TeV}}$.

The small mixing make electroweak precision constraints (which can be
straightforwardly generalised to observables beyond $S,T,U$ and
flavour constraints in the present model) redundant: The region
excluded by the current $S,T$ ellipse corresponds to large mixing
$\lambda_{\rm{P}} \gtrsim 10^2$, a region which is well excluded by RG
arguments. In this sense, electroweak precision does not yield an
additional constraint, but is implied by the consistency of the model
itself.

The suppression of single-${h_2}$ phenomenology can in principle be
counteracted in the di-Higgs channels $pp\to {h_2}\to h_1h_1\to
{\text{SM}}$ (see {\it e.g.} Ref.~\cite{dien} for an example in a
different context). The resonance is extremely narrow, and for the
parameter space $m_{h_2}>2m_{h_1}$ it naturally appears in the TeV
regime. Such signatures have been investigated in
\cite{bowen,Dolan:2012ac}. While the small mixing angle naively means
a suppressed $s$-channel contribution of ${h_2}$ to the di-Higgs
phenomenology, it exclusively decays to a SM-like di-Higgs system in
our setting with a potentially large coupling $\sim \lambda_{\text{P}}
v \sim 1$~GeV. The small coupling of ${h_2}$ to the top quarks running
in the gluon fusion loops however can typically not be beaten by the
${h_2} h_1h_1$ vertex. This contribution has to be put in contrast to
the off-shell $h_1h_1h_1$ vertex $\sim v \gg \lambda_{\text{P}}
\left\langle |\phi|\right\rangle$ which is SM-like and, more
importantly for high energetic Higgses, to the box-induced continuum
$gg\to h_1h_1$ production. We have performed a full one-loop
computation of $pp\to h_1h_1\to \{\text{visible}\}$ via gluon fusion
(which by far the most dominant production mode in the SM) in the
proposed model and have scanned the cross section for a couple of
parameter points and always find a di-Higgs cross section of
${\cal{O}}(16)$~fb. This agrees with the tree-level SM result
\cite{smres} within uncertainties and we expect that adapting SM
Higgs-like searches for the heavy $h_2$ is going to result in more
solid constraints earlier.

In total, precision analyses of the Higgs-like candidate at 125~GeV
and extending Higgs boson-like searches beyond $1~{\text{TeV}}$
therefore provide the best handles to constrain this model in its
simplest implementation. The portal parameter, which is required to be
small in the limit of light $h_2$ can be efficiently constrained by
measuring the $h_1$ couplings at a future linear collider. Excluding
heavy $h_2$ fields in high luminosity LHC searches limit the parameter
space for $\lambda_{\rm{P}}\gtrsim 0.001$.

Low energy measurements on the other hand are highly sensitive to very
light masses, {\it{e.g.}} fifth force measurements can probe mixing
angles $\sin\vartheta<10^{-10}$ for $m_{h_2}\lesssim
10^{-2}~\text{eV}$~\cite{Ahlers:2008qc}, which limits the model for
such very small $(\lambda_{\rm P},m_{h_2})$ combinations.  For
moderate masses $m_{h_{2}}\lesssim 100~{\rm keV}$ stellar evolution
sets strong constraints on scalar couplings to two
photons~\cite{stellar}.  The coupling of $h_{2}$ to two photons is
given by,
\begin{equation}
g_{h_{2}\gamma\gamma}=\sin(\vartheta)g^{\rm SM}_{h\gamma\gamma},
\end{equation}
we can translate these bounds into a limit on $\sin(\vartheta)\lesssim
10^{-3.86}$ for masses $m_{h_{2}}\lesssim 100~{\rm keV}$.

\section{Conclusions}
\label{sec:five}

The Coleman-Weinberg mechanism is an intriguing possibility to
naturally generate a very small scale.  However, if done within the
Standard Model its main prediction of a very light Higgs (far below
the $Z$-mass) is in clear conflict with the experimental observation
of a Higgs(-like) particle at $\sim$125~GeV. In this paper we have
shown that a simple Higgs-portal model allows to generate the
electroweak scale via the Coleman-Weinberg mechanism while at the same
time giving a phenomenologically viable Higgs mass.  The simple model
we have discussed has a rich phenomenology and can be tested at the
LHC.

While the explicit model considered in this paper is
interesting on its own right thanks to its simplicity, it can also be
viewed as a representative of a whole class of models in which the
Coleman-Weinberg mechanism generates a low scale in the hidden sector
which then is transmitted to the SM via the Higgs
portal.

An essential requirement for the Coleman-Weinberg mechanism to work is
that the renormalised mass at the origin of the potential
vanishes. All scales are then generated from dimensional transmutation
and are exponentially suppressed compared to the UV scale at which the
theory is initialised. We have collected and discussed various
arguments why the vanishing of the renormalised mass terms is a
sensible condition.  We find two main possibilities.
\begin{enumerate}
\item Let us take the {\it full} classical theory to be massless and
  scale invariant. Scale invariance is broken in the quantum
  theory. Dimensional regularisation is the scheme which (to our
  knowledge) breaks scale invariance minimally. In dimensional
  regularisation the condition of vanishing masses at the origin is
  independent of the renormalisation scale and can therefore be
  imposed consistently without fine-tuning (in a more general scheme a
  similar condition can be defined consistently). The radiative
  generation of the EWSB scale in the full theory then proceeds via
  the CW mechanism as described in the body of the paper.
\item Alternatively, assume that only the {\it low energy
    theory} we observe, has approximate scale invariance
  up to quantum corrections. The scale invariance breaking effects of
  additional high scale physics cancels exactly (not approximately as
  one would require to generate a small renormalised mass scale at the
  origin of field space).
\end{enumerate}

The model's phenomenology is that of a Higgs portal model, however
with constraints imposed that arise from generating the electroweak
scale via a small visible-hidden sector coupling. The modifications
compared to the SM are generically small, and exclusion bounds are
driven by precision investigations of the Higgs boson candidate. In
essence, electroweak symmetry breaking proceeds along the lines of the
SM, with modifications only due to small mixing effects and total
Higgs width modifications. All these quantities can be determined most
precisely at a future linear collider.

Although our simple setup cannot be considered a full solution to the
hierarchy problem it provides a simple and experimentally testable
scenario that may act as a first step to gain additional insight on
the mechanism that generates the electroweak scale.

\section*{Acknowledgements}
We would like to thank A. Hebecker, J.~Pawlowski, J.~Redondo,
T. Plehn, B.~Stech and C. Wetterich for helpful and stimulating
discussions.
CE acknowledges funding by the Durham International Junior Research
Fellowship scheme. VVK gratefully acknowledges the support of the
Wolfson Foundation and the Royal Society.

\bibliographystyle{h-physrev5}

\end{document}